\documentclass[a4paper]{PoS}

\usepackage{myjournal}
\usepackage{atlasphysics}

\usepackage{floatrow}
\title{ATLAS SM VH(bb) Run-2 Search}

\ShortTitle{ATLAS SM VH(bb) Run-2 Search}

\author{\speaker{Adrian BUZATU\thanks{On behalf of the ATLAS Collaboration.}}\\
        University of Glasgow\\
        E-mail: \email{adrian.buzatu@glasgow.ac.uk}}

\abstract{}

\abstract{The Higgs boson discovered at the LHC in 2012 has been observed coupling directly to $W$ and $Z$ bosons and to $\tau$ leptons, and indirectly to top quarks. In order to probe whether it is indeed the particle predicted by the Standard Model, direct couplings of the Higgs boson to quarks must also be measured. The Higgs boson decays most often to a pair of bottom quarks (with a branching ratio of 58\%). When the Higgs boson is produced alone in gluon-gluon fusion, the signal in this decay mode is overwhelmed by the regular multi-jet background. By requiring the Higgs boson to be produced in association with a vector boson $V$ ($W$ or $Z$), which is further required to decay leptonically, data events can be selected using charged-lepton or missing transverse energy triggers. The Tevatron experiments presented combined results showing evidence for the \vh~process at a significance level of about 3 standard deviations, while the combined LHC results from Run-1 data show a 2.6 standard deviation evidence for the \hbb~decay mode. In this poster, the ATLAS \vh~search using Run-2 data is summarised.}

\FullConference{38th International Conference on High Energy Physics\\
		3-10 August 2016\\
		Chicago, USA}

\begin{document}

\section{Introduction}

A Higgs boson scalar particle with a mass of 125 GeV has been discovered by ATLAS~\cite{ATLASDetector} and CMS using the LHC data. The Standard Model of elementary particles and their interactions (SM) is known not to explain key questions such as baryogenesis, the nature of dark matter, and force unification. It is believed that the SM is a low energy approximation of a larger theory of new physics processes denoted beyond the SM, or BSM. It is important to measure precisely the coupling of the 125 GeV boson to the SM particles to find out whether it is a SM or a BSM Higgs boson. The boson decays to gauge bosons and $\tau$ leptons have been observed, but not to quarks. The largest quark decay branching ratio is to bottom quark pairs (\hbb), which is 58\% in the SM. If the Higgs boson is produced alone in gluon fusion, the multi-jet QCD background is overwhelming. To improve the signal to background ratio, the associated production of a Higgs boson and a vector boson ($VH$) is employed. The ATLAS 8 TeV search from Run-1 measured a signal strength of $\mu = 0.51 \pm 0.40 {\rm (stat.)} ^{+0.31}_{-0.25} {\rm (syst.)}$~\cite{ATLASRunI8TeVPaperVH}, consistent with the SM, with the statistical error dominating. In this poster proceeding the first ATLAS \vh~Run-2 conference note~\cite{ATLASRunII2016ICHEPVH} result is summarised. 

\section{Analysis strategy}

There are three signal processes depending on the number of charged leptons: 0, 1 or 2. The dominant backgrounds are \ttbar~(reducible with improved $b$-tagging) and $V+\bbbar$ (irreducible). 

\begin{figure}
\includegraphics[height=2.2cm]{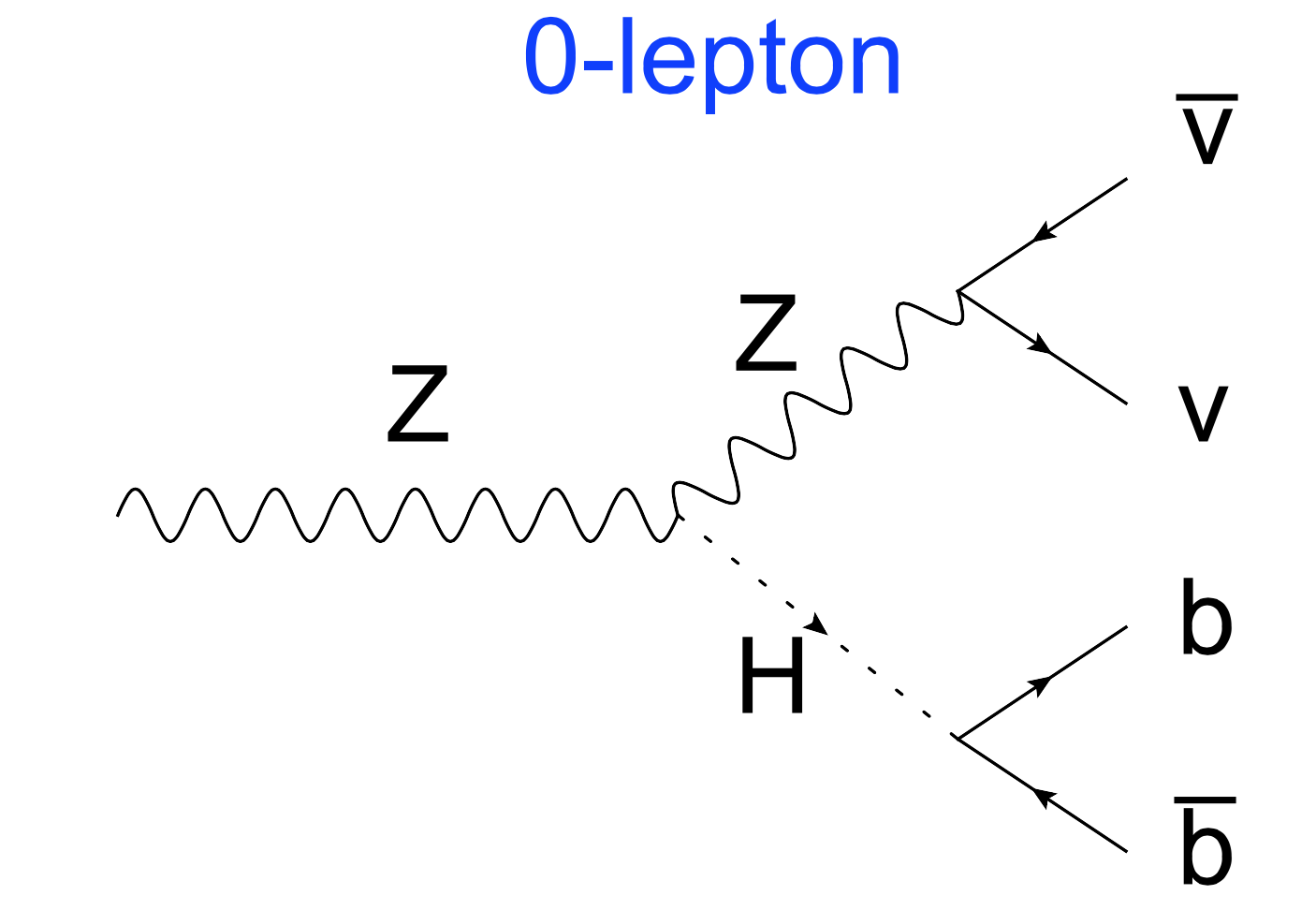}
\includegraphics[height=2.2cm]{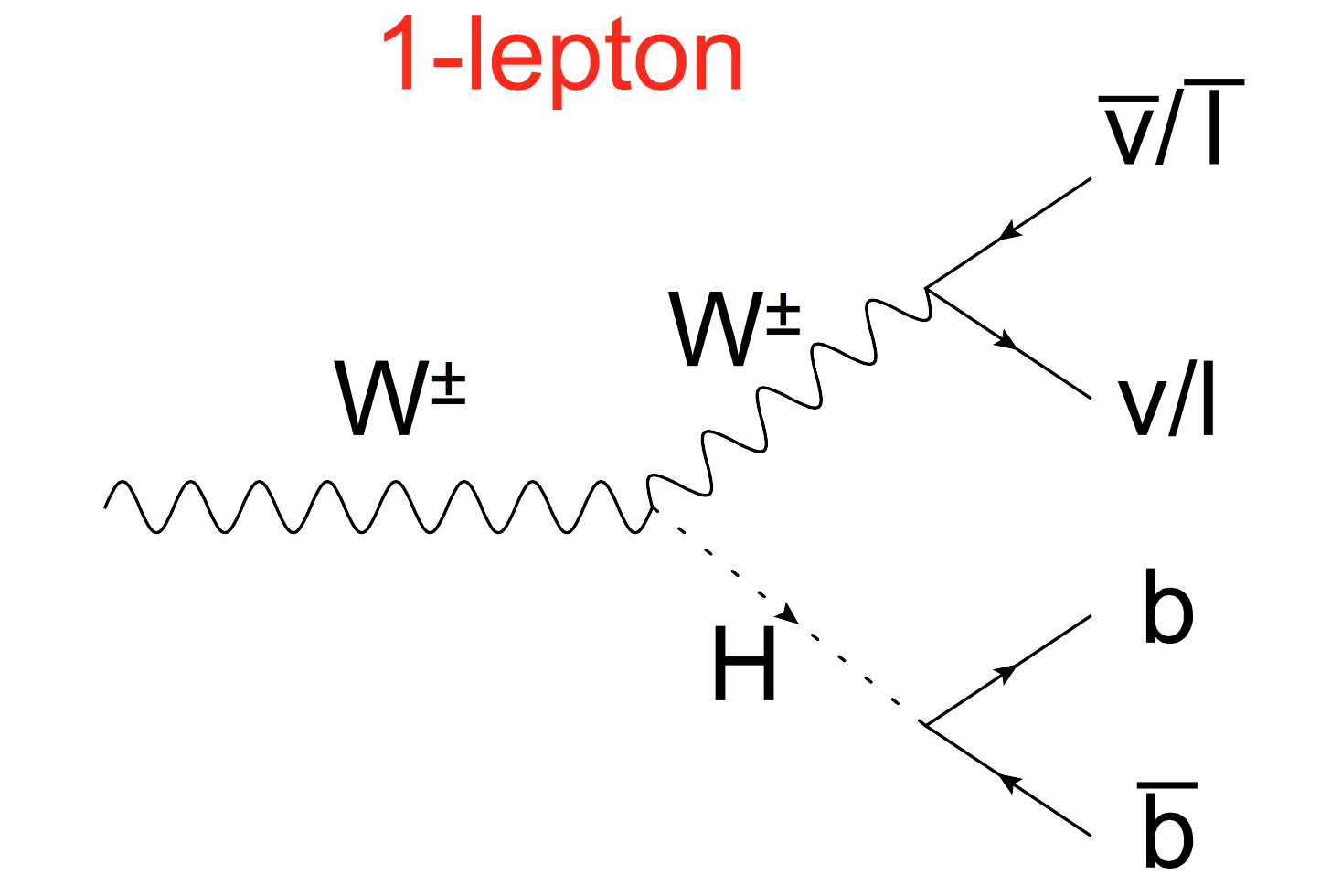}
\includegraphics[height=2.2cm]{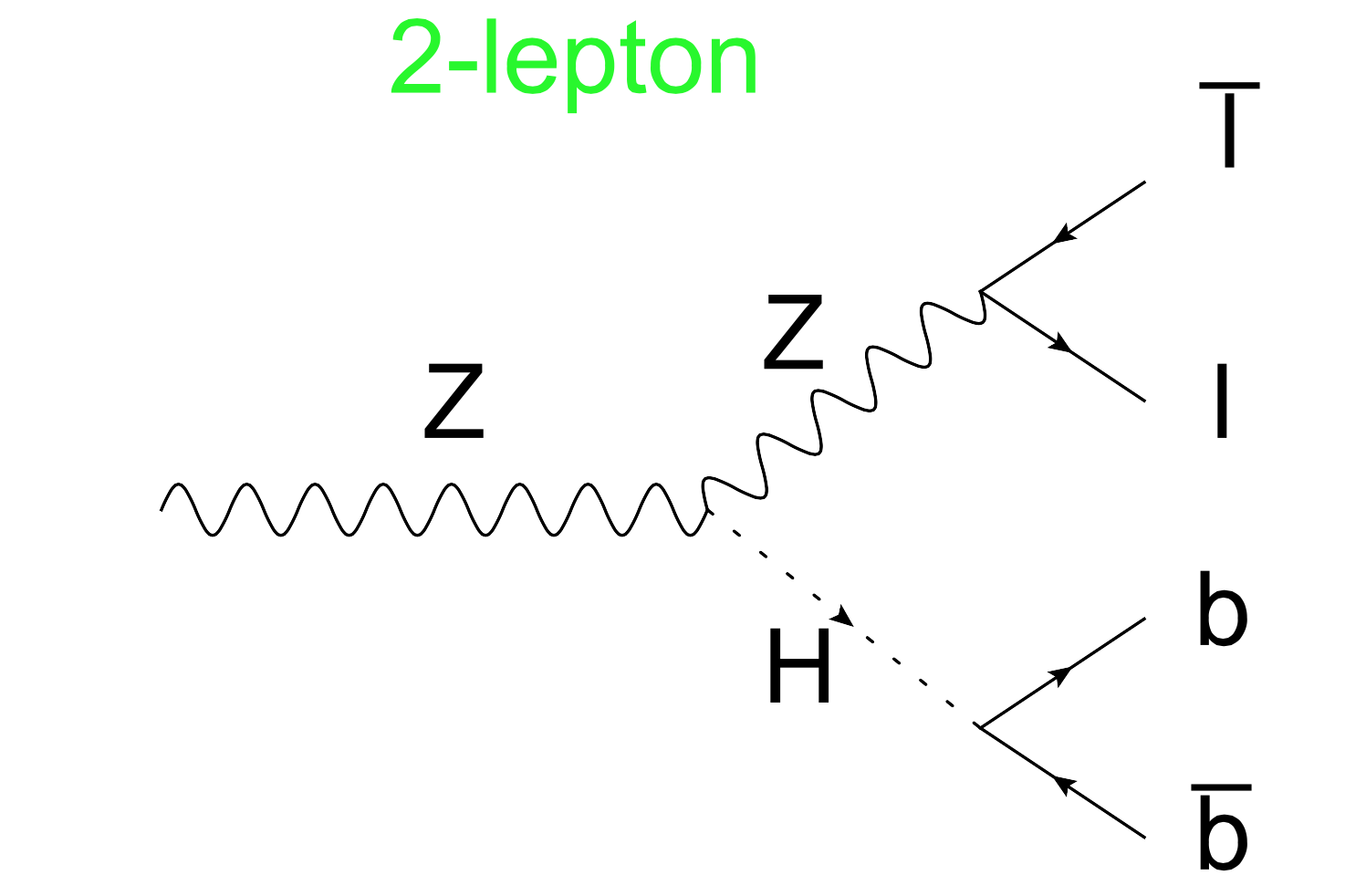}
\includegraphics[height=2.2cm]{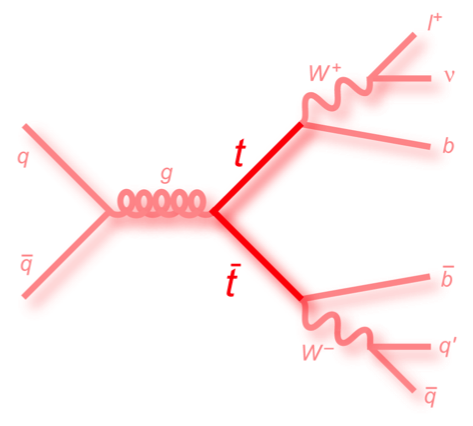}
\includegraphics[height=2.2cm]{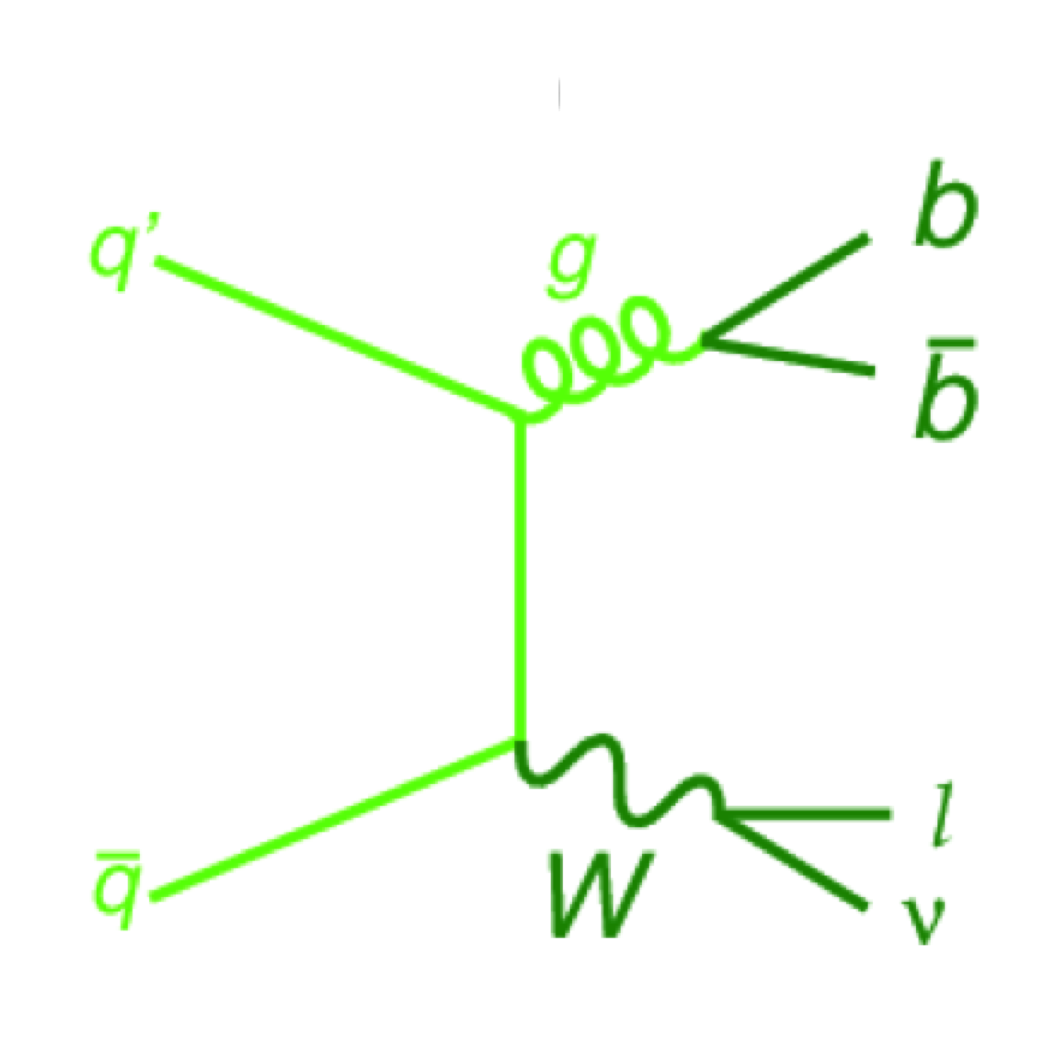}
\caption{The three signal processes and the \ttbar~and $W+\bbbar$ dominant background processes.}
\label{fig:signals}
\end{figure}

To improve sensitivity, the kinematic phase space is split by number of charged leptons, jets, $b$-jets, and the transverse momentum of the vector boson $V$ denoted \ptv, as described in Table~\ref{tab:categories}.

\begin{table}[b]
\begin{center}
\begin{tabular}{|l|l|l|l|}
\hline
- & 0-lepton & 1-lepton & 2-lepton \\
\hline
Number of jets & 2, 3 & 2, 3 & 2, $\ge$3 \\
Number of $b$-tags & 2 & 2 & 2 \\
\ptv & $\ge$ 150 GeV &  $\ge$ 150 GeV &  $\ge$ 150 GeV \\
\ptv & - &  - &  $\le$ 150 GeV \\
\hline
\end{tabular}
\end{center}
\caption{Event selection cuts for the different analysis categories~\cite{ATLASRunII2016ICHEPVH}.}
\label{tab:categories}
\end{table}

\begin{figure}[b]
\includegraphics[width=0.30\textwidth]{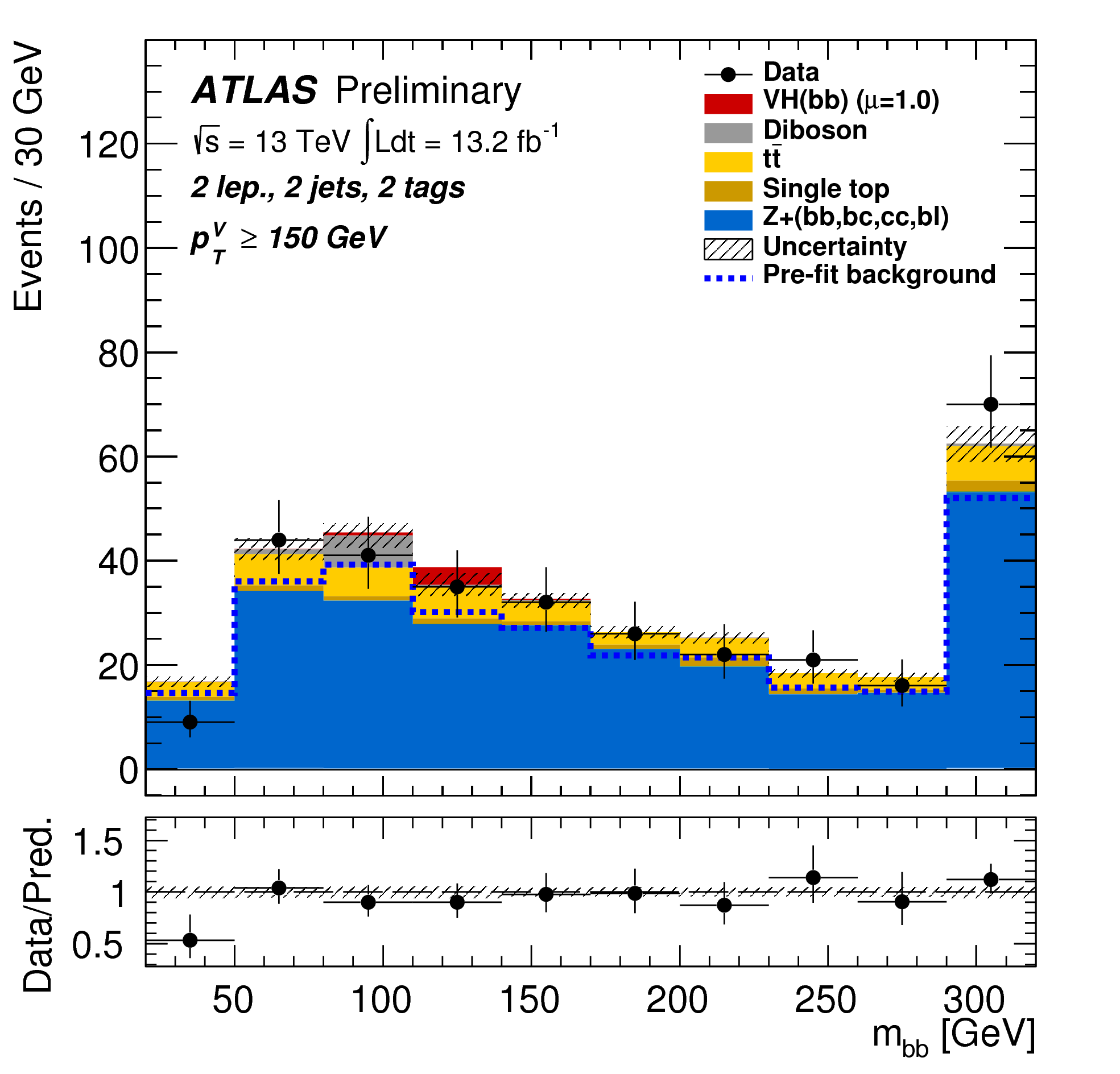}
\includegraphics[width=0.30\textwidth]{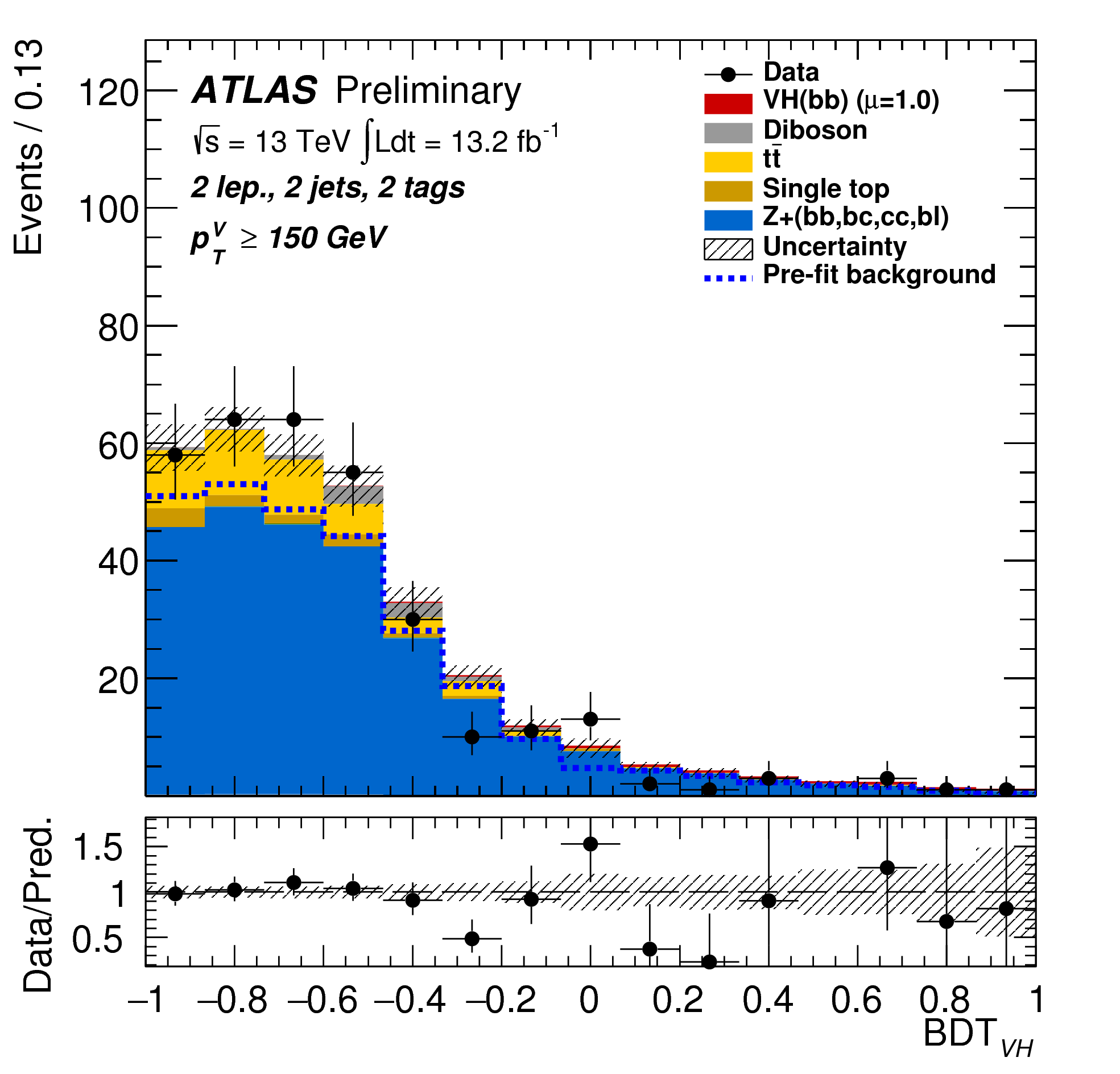}
\includegraphics[width=0.38\textwidth]{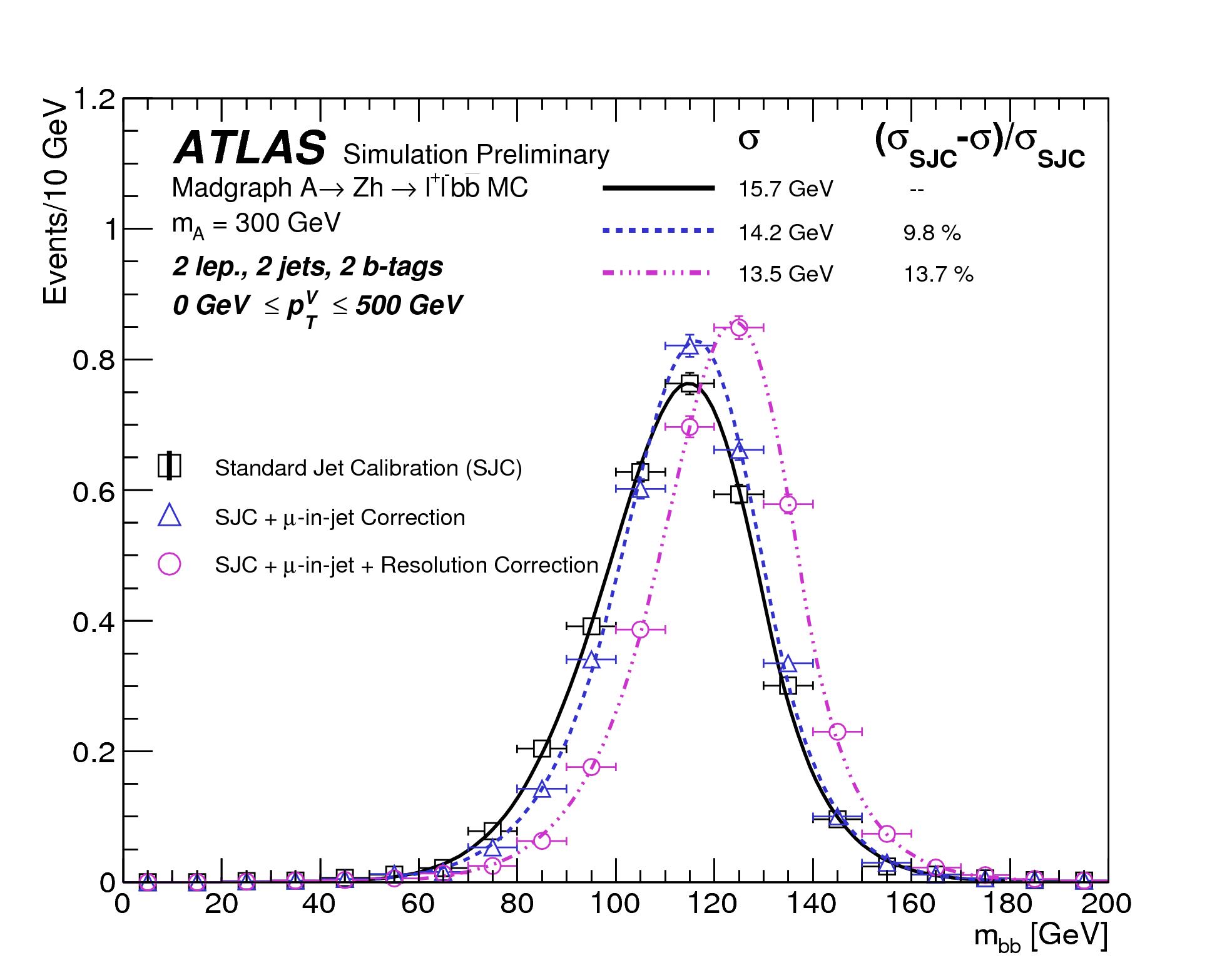}
\caption{\mbb~(left) and BDT (centre) output for the most sensitive category: 2-lepton, 2 jets, 2 $b$-jets, $\ptv > 150$ GeV~\cite{ATLASRunII2016ICHEPVH}. The signal \mbb~distribution after the $b$-jet energy corrections in a \azh~signal (right)~\cite{ATLASRunII2016MoriondAZh}.}
\label{fig:kinematic}
\end{figure}

The  best signal (S) to background (B) discriminant is the di-$b$-jet invariant mass (\mbb). The \vh~signal is peaked around 125 GeV, while the $Z$+jets and \ttbar~have a wide distribution, as shown in Figure~\ref{fig:kinematic} (left). To discriminate further, we use a multivariate technique by training a boosted decision tree (BDT). It uses as inputs \mbb~and additional kinematic event variables. The BDT output in Figure~\ref{fig:kinematic} (centre) peaks to the left for background and to the right for signal.

\section{$b$-jet energy corrections}

\begin{figure}
\includegraphics[height=4.3cm]{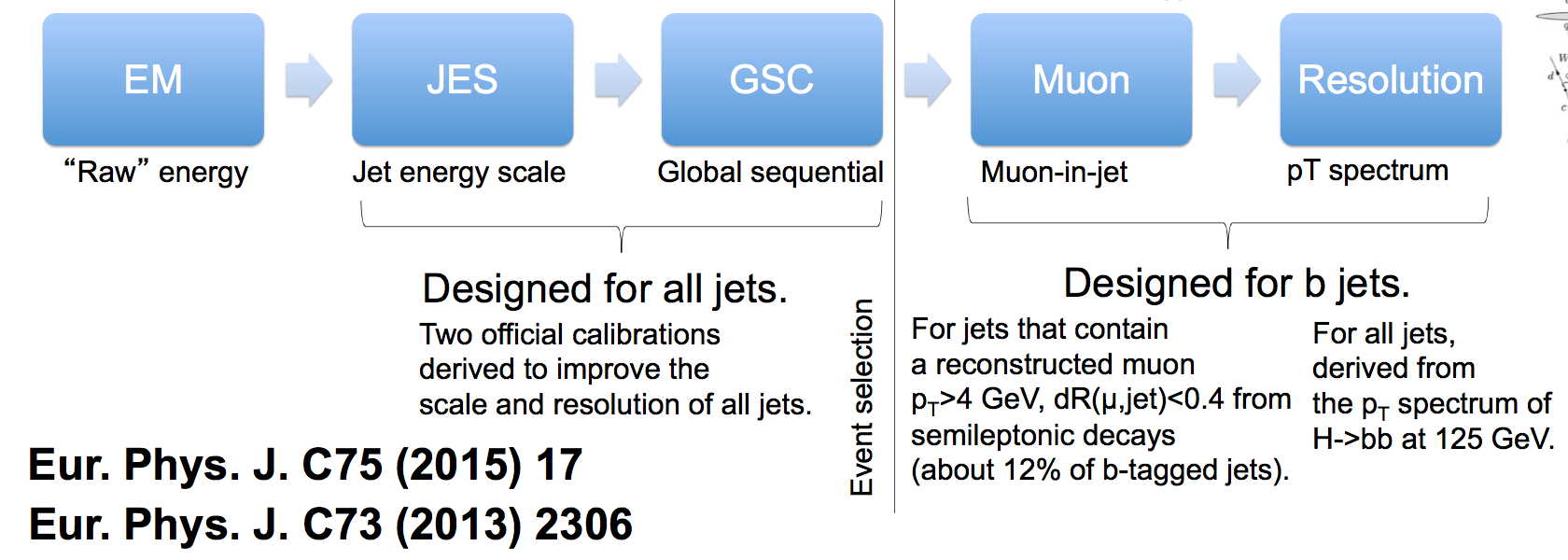}
\caption{Sequence of jet energy corrections for ATLAS $b$-jets from \hbb.}
\label{fig:bJetCorr}
\end{figure}

The \mbb~reconstruction is improved if the energy calibration of each of the two $b$-tagged jets is improved. ATLAS jets are reconstructed with an anti-$k_t$ algorithm with a cone radius $R=0.4$. ATLAS applies a sequence of four energy calibrations illustrated in Figure~\ref{fig:bJetCorr}. The former two are applied for all the jets, and the latter two are $b$-jet specific. The first correction, denoted jet energy scale (JES)~\cite{ATLASRunI7TeVJERGSC}, corrects for several effects: primary vertex position, pile-up area subtraction, as well as the jet energy scale corrections. The second correction, global sequential calibration (GSC)~\cite{ATLASRunI7TeVJERGSC}, improves the jet energy resolution. After this stage, the jet and event selection is made: two $b$-tagged jets with $\pt > 20$ GeV, one of which has $\pt > 45$ GeV. 

Additional corrections are needed for $b$-tagged jets. The $\mu$-in-jet correction is for the 12\% of $b$-jets that contain a reconstructed muon from a $B$ hadron semileptonic decay (SLD). The jet 4-vector measured in the calorimeter has the $\mu$ 4-vector measured in the inner detector and $\mu$-spectrometer added, while the 4-vector of the energy deposited in the calorimeter by the $\mu$ is removed. In the PtReco correction, the jet 4-vector thus obtained is multiplied by a factor, which is determined as a function of jet \pt~and whether there is a SLD. It is the mean of the inverse of the response of the jet \pt~relative to the particle-level-jet including $\mu$s and $\nu$s from SLD. It accounts for out-of-cone energy deposition due to large impact parameter tracks, and for the $\nu$ from SLD. With each $b$-jet energy correction, the peak and width of \mbb~are improved, as seen in Figure~\ref{fig:kinematic} (right). 
\section{Simultaneous fit, validation and results}

\begin{figure}
\includegraphics[height=3.8cm]{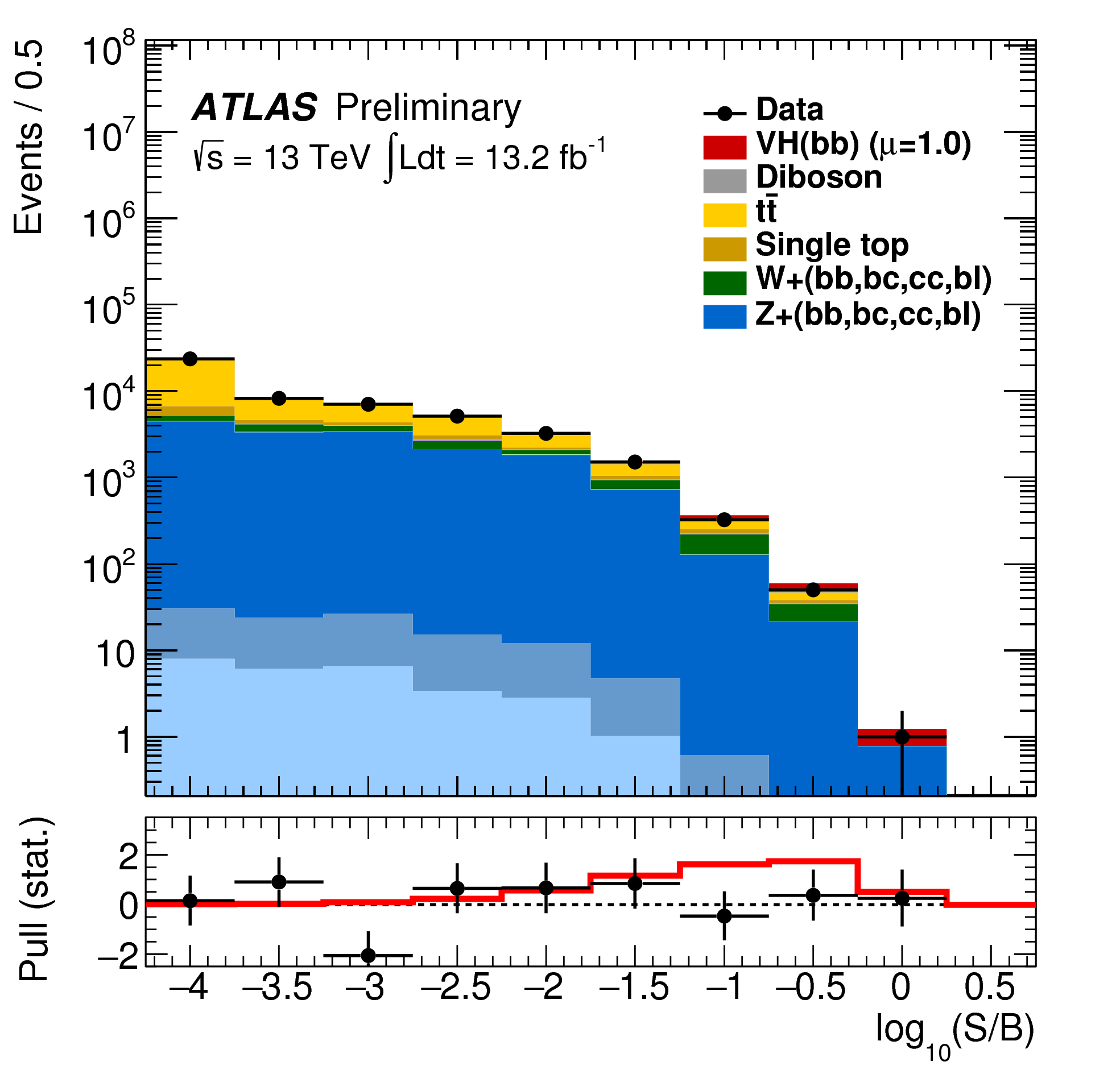}
\includegraphics[height=3.8cm]{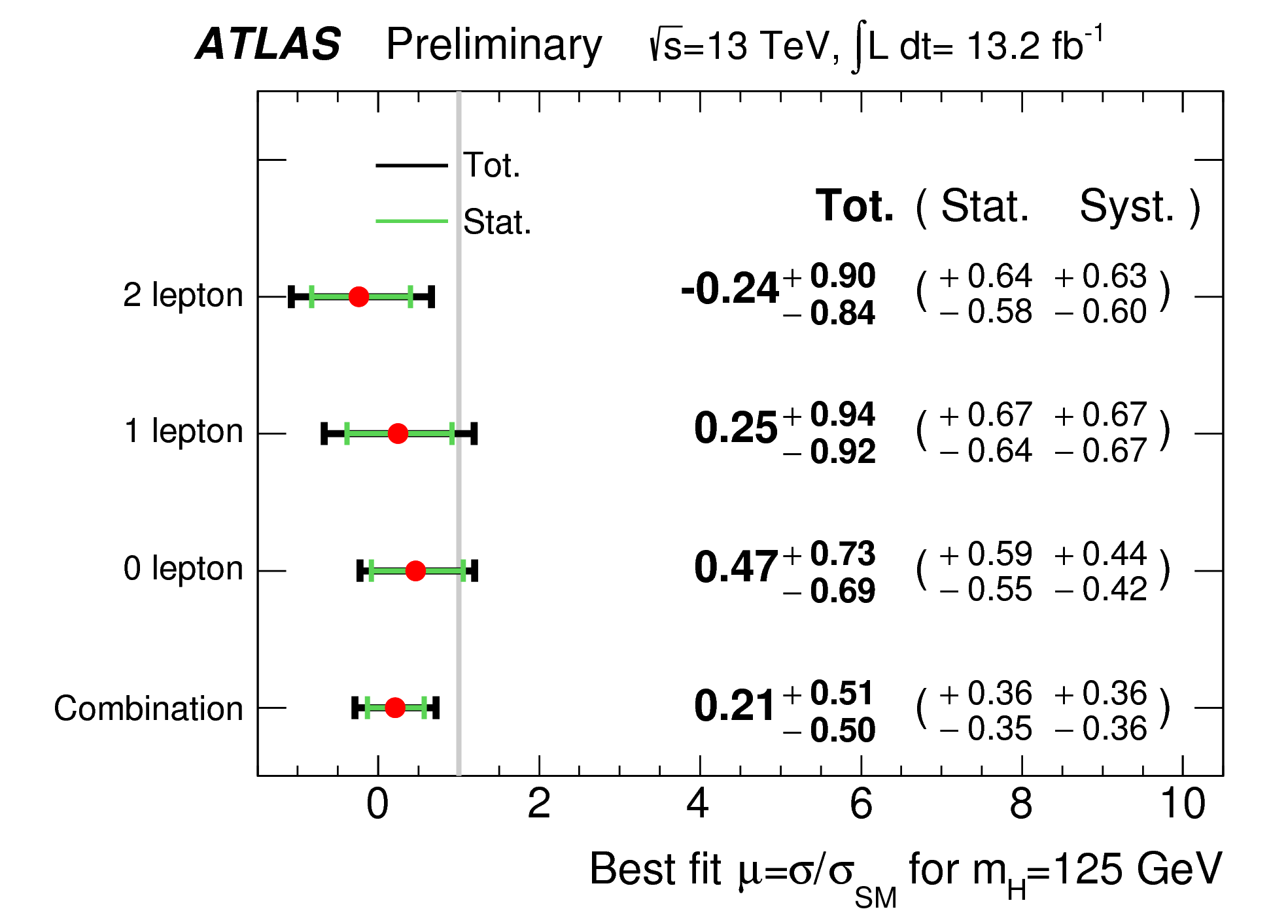}
\includegraphics[height=3.8cm]{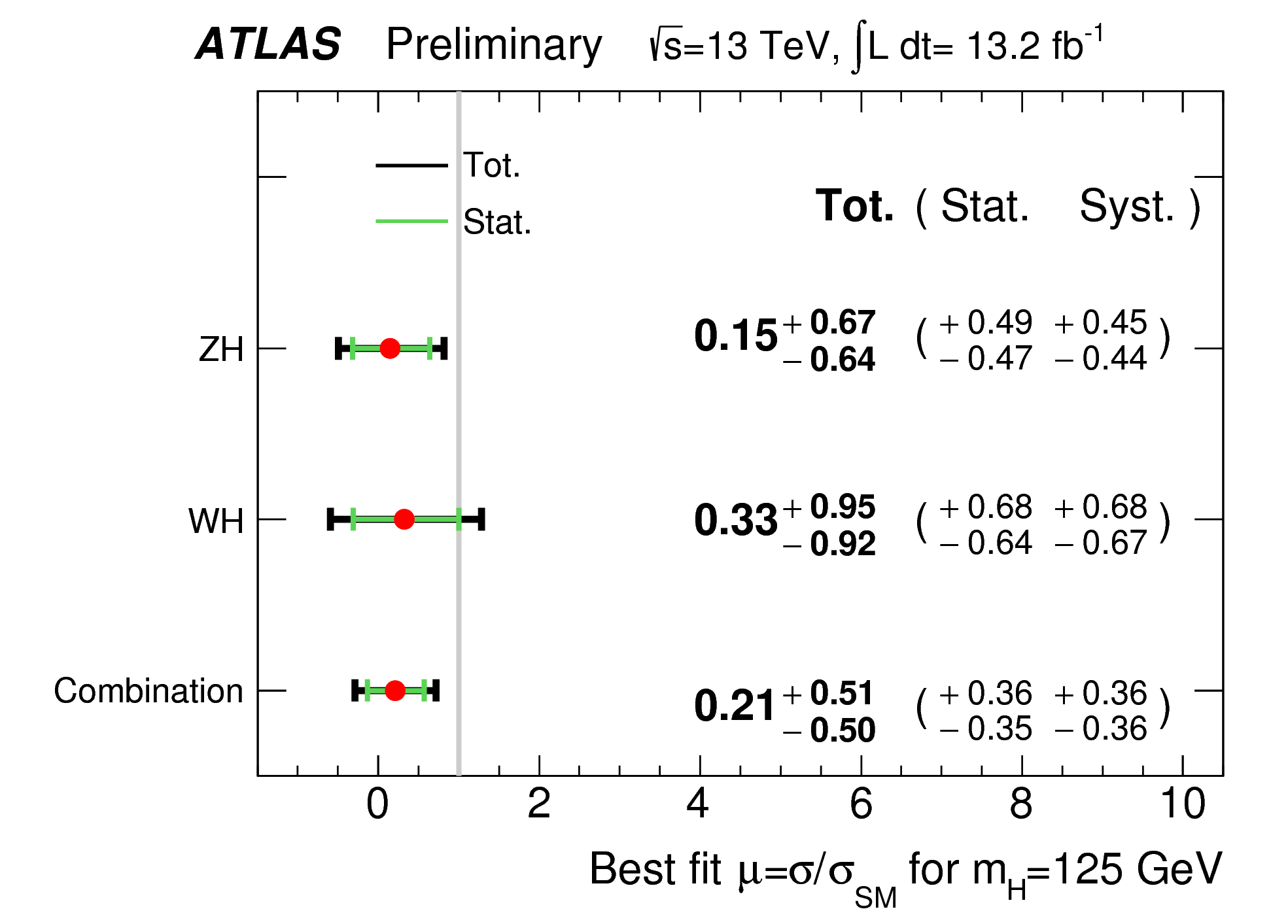}
\caption{All signal categories combined into log(S/B) bins (left). The lower panel shows the pull of the data with respect to the B-only prediction using the statistical uncertainty only. The fitted values of the measured S strength broken down into the charged lepton categories (centre) and the production channels (right)~\cite{ATLASRunII2016ICHEPVH}.}
\label{fig:result}
\end{figure}

To constrain the backgrounds and get a limit and signal strength, we perform a simultaneous fit over all categories. Systematic uncertainties are treated as nuisance parameters. The signal strength, and the background normalisation of $V$ + heavy flavor are floated in the fit. The procedure is validated by searching for the known process  $VZ\rightarrow V\bbbar$. The BDTs are trained with $VZ$ as signal and \vh~as background. The measured signal strength is $\mu_{VZ}~= ~0.91 ~\pm~0.17 ~{\rm~(stat.)}~^{+0.32}_{-0.27} ~{\rm~(syst.)}$. The observed (expected) significance is of 3.0 (3.2) standard deviations ($\sigma$). Given this validation, the BDTs were trained with \vh~as signal and $VZ$ as background. With the background normalisations from the fit values, the $S/\sqrt(B)$ plot summing all categories is shown in Figure~\ref{fig:result} (left). The best signal strength is $\mu=0.21 \pm 0.36 {\rm (stat.)} \pm {\rm 0.36 (syst.)}$. The values split by the number of charged leptons and by signal processes are shown in Figure~\ref{fig:result} (middle and right).

\section{Conclusions}
The measured signal strength is $\mu=0.21 \pm 0.51$, consistent with the SM within 2$\sigma$. The relative contribution of the systematic uncertainty has increased with respect to the Run-1 result. It contributes now as much as the statistical one. 

\bibliographystyle{JHEP}
\bibliography{BuzatuProceedingICHEP2016} 

\end{document}